\documentclass[%
superscriptaddress,
A4,
amsmath,amssymb,
aps,
prl,
twocolumn,
longbibliography,
]{revtex4}

\usepackage{graphicx}% Include figure files
\usepackage{dcolumn}% Align table columns on decimal point
\usepackage{bm}% bold math
\usepackage{xcolor}
\usepackage{float}
\usepackage{verbatim}
\begin{document}

%\preprint{APS/123-QED}

\title{CPA-Lasing in Thin-Elastic Plates via Exceptional Points}% Force line breaks with \\

\author{M. Farhat}
\email{mohamed.farhat@kaust.edu.sa}
\affiliation{Computer, Electrical, and Mathematical Science and Engineering (CEMSE) Division, King Abdullah University of Science and Technology (KAUST), Thuwal 23955-6900, Saudi Arabia}
 
\author{P.-Y. Chen}%
\affiliation{Department of Electrical and Computer Engineering, University of Illinois at Chicago, Chicago, Illinois 60607, USA}

\author{S. Guenneau}
\affiliation{UMI 2004 Abraham de Moivre-CNRS, Imperial College London, London SW7 2AZ, United Kingdom}

\author{Y. Wu}
\email{ying.wu@kaust.edu.sa}
\affiliation{Computer, Electrical, and Mathematical Science and Engineering (CEMSE) Division, King Abdullah University of Science and Technology (KAUST), Thuwal 23955-6900, Saudi Arabia}

\begin{abstract}
%Achieving extraordinary amplification and lasing in acousto-elastic systems is a long-sought for goal. We present here how a coherent perfect absorber-laser (CPAL) enabled by parity-time (PT) symmetry breaking may be exploited to build monochromatic amplifying devices for flexural waves. To this end, we apply the transfer ($M$) matrix method and we analyze the scattering ($S$) matrix formalisms for flexural waves propagating in thin-plates, a system akin to Fabry-Perot, as well as develop $\mathcal{PT}$-symmetry theory and its applications to thin-elastic plates, known to support both propagating and evanescent waves. The fourth order partial differential equation governing the propagation of flexural waves leads to four by four $M$ and $S$ matrices, and this results in physical properties of the $\mathcal{PT}$-symmetry specific to elastic plate systems. We thus demonstrate the possibility of using coherent perfect absorber lasing for such systems and we argue the possibility of using such CPAL device to detect extremely small-scale vibration perturbations with important outcomes in surface science (imaging of nanometer vibration) and geophysics (improving seismic sensors like velocimeters). The device can also generate finite signals using very low exciting intensities. The system can alternatively be used as a perfect absorber for flexural energy by tailoring the left and right incident energy with many applications ranging from surface science to civil engineering.

We present here how a coherent perfect absorber-laser (CPAL) enabled by parity-time ($\mathcal{PT}$)-symmetry breaking may be exploited to build monochromatic amplifying devices for flexural waves. The fourth order partial differential equation governing the propagation of flexural waves leads to four by four transfer matrices, and this results in physical properties of the $\mathcal{PT}$-symmetry specific to elastic plate systems. We thus demonstrate the possibility of using CPAL for such systems and we argue the possibility of using this concept to detect extremely small-scale vibration perturbations with important outcomes in surface science (imaging of nanometer vibration) and geophysics (improving seismic sensors like velocimeters). The device can also generate finite signals using very low exciting intensities. The system can alternatively be used as a perfect absorber for flexural energy by tailoring the left and right incident wave for energy harvesting applications.
\end{abstract}

\date{\today}% It is always \today, today,

\maketitle


%\tableofcontents
%%%%%%%%%%%%%%%%%%%%%%%%%%%%%%%%%%%%%%%%%%%%%%%%%%%%%%%%%%%%%%%%%%%%%%%%%%%%%%%%%%%%%%%%%%%%%%%%%%%
%\textit{Introduction.}---
In recent years, the use of resonant elements enriched the properties of periodic media, with the paradigm shift of metamaterials. These are constructed from a judicious arrangement of physical resonators whose size is very small compared to the typical wavelength of interest \cite{pendry1999magnetism} and permit some exotic applications such as negative refraction \cite{pendry2000negative,smith2004metamaterials} or scattering cancellation technique (SCT) \cite{alu2005achieving,chen2012invisibility}. Several research groups have worked on the extension of metamaterials and metasurfaces to elastic waves in solid structures \cite{movchan2007bloch,wu2011elastic}. For instance, the tensorial nature of the equations governing elastic waves requires complex analytical and numerical modeling that takes into account the coupling between pressure and shear waves at solid interfaces \cite{timoshenko1959theory}. 
%Hence, the control of these waves through elastic metasurfaces, may pave the way to important applications in civil engineering, such as seismic cloaking \cite{brule2020emergence}, by reducing or guiding the destructive waves around sensitive structures, or the reduction of vibration in sensitive components of cars or airplanes \cite{ungar1990vibration}. 
In the same vein, a particular type of elastic solid, the thin elastic plate (TEP), has drawn a growing interest in the wave physics community \cite{timoshenko1959theory,norris1995scattering}. The plate has a small vertical dimension (thickness) in comparison to its lateral dimensions and the wavelength \cite{timoshenko1959theory}, resulting in the vertical displacement of the plate largely determined by the flexural mode (i.e. no shear), sometimes designed as $A_0$ mode \cite{timoshenko1959theory}. The bending of these TEPs can be described by the Kirchhoff-Love equation (fourth order partial differential equation (PDE)) and interestingly has a scalar nature in the case of isotropic plates \cite{timoshenko1959theory}. This feature allows for a more straightforward numerical modeling of waves propagating in isotropic TEPs. Subsequently, several designs have been proposed for flexural waves, including cloaking \cite{farhat2009ultrabroadband,zhu2018elastic}, negative refraction \cite{dubois2014flat}, localized surface plate modes \cite{farhat2017localized}, SCT \cite{farhat2014platonic}, elastic plate crystals \cite{mcphedran2015parabolic}, etc.

On another side, it was shown in 1998 that non-Hermitian Hamiltonians with Parity-Time ($\mathcal{PT}$)-symmetry have real eigenvalues \cite{bender1998real}. First used in quantum mechanics \cite{mostafazadeh2002pseudo}, this feature was then applied to optics because the paraxial wave equation is mathematically equivalent to the Schr\"{o}dinger equation \cite{makris2008beam,ruter2010observation}, leading to some remarkable properties, such as an asymmetric propagation of the modes or the existence of an exceptional point (EP) where the $\mathcal{PT}$-symmetry is broken \cite{feng2017non,sakhdari2017pt}. $\mathcal{PT}$-symmetry gained a tremendous momentum among the photonics community due to its promising outcomes, e.g. environmental sensing \cite{chen2016p}, on-chip optical systems \cite{peng2014parity}, cavity-mode selection in microring lasers \cite{feng2014single}. In the same vein, it was shown that acoustic waves exhibit such non-reciprocal behavior when loss and gain layers are balanced \cite{zhu2014p,christensen2016parity}. Hence, although these $\mathcal{PT}$-symmetric acoustic systems are still at an early stage, several promising applications have been recently envisioned, e.g. unidirectional invisibility cloaking \cite{li2019ultrathin}, invisible acoustic sensor \cite{fleury2015invisible}, phononic laser \cite{zhang2018phonon}, and acoustic Willis coupling \cite{quan2019nonreciprocal}. 
%For these designs, and in order to overcome the lack of acoustic gain in the available materials found in nature, an active sound generation unit is employed, requiring some complex circuitry and an external supply of power. So-called acoustic Willis coupling has been also invoked to realize some non-reciprocal features \cite{quan2019nonreciprocal}. 
With regards to elastodynamics waves, shunted piezoelectric thin materials may lead to gain/loss in elastic plates, depending on the resistance of the shunted circuit \cite{vasseur2007waveguiding,hladky1993finite,hou2018tunable}, which was previously used to realize negative refraction \cite{hou2018p}. Flexural waves in beams were further shown to possess $\mathcal{PT}$-symmetric effects \cite{wu2019asymmetric}. In Ref. \cite{lu2020non}, a different technique was employed to produce non-reciprocal wave transmission.

%\textcolor{red}{We propose in this Letter, to use the transfer matrix ($M$) and scattering matrix ($S$) formalisms to rigorously analyze scattering of flexural waves in TEPs with alternating gain and loss layers. We note the possibility of having asymmetric reflectionless behavior, when gain and loss are balanced. We further show that this effect takes place around the EP of such devices. Quite interestingly, we show that it is possible to tune the frequency of the EPs quite largely and in different ways, due to the peculiar dispersion relation of flexural waves in thin-plates.} 

We show in this Letter the possibility to realize the equivalent of lasing in elastic plates, i.e., coherent perfect absorber laser (CPAL) thanks to gain and loss values corresponding to the lasing threshold displaying a quantized behavior, which occurs due to topological character of the system. The spectral singularity could be also used for coherent perfect absorber in elastic plates.

%%%%%%%%%%%%%%%%%%%%%%%%%%%%%%%%%%%%%%%%%%%%%%%%%%%%%%%%%%%%%%%%%%%%%%%%%%%%%%%%%%%%%%%%%%%%%%%%%%%
%%%%%%%%%%%%%%%%%%%%%%%%%%%%%%%%%%%%%%%%%%%%%%%%%%%%%%%%%%%%%%%%%%%%%%%%%%%%%%%%%%%%%%%%%%%%%%%%%%%
%\label{sec:section2}
%\subsection{Biharmonic equation and transfer matrix}
\begin{figure*}[t!]
    \centering
    \includegraphics[width=1.45\columnwidth]{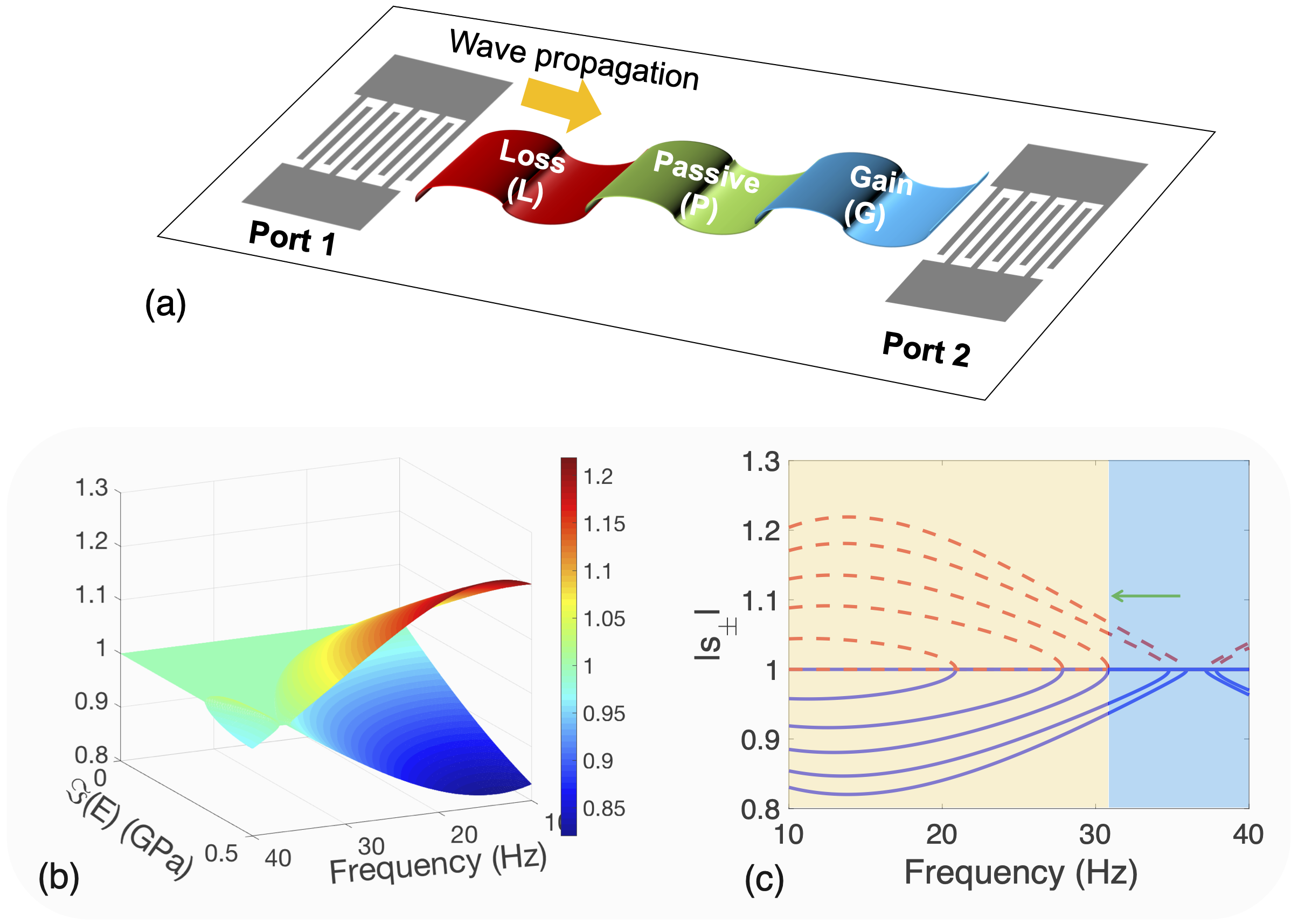}
    \caption{(a) Structure of the gain/loss device. (b) 2D plot of the eigenvalues versus frequency and the imaginary part of the Young modulus. (c) Eigenvalues for specific values of the loss/gain in the Young modulus (0, 0.1, 0.2, 0.25, 0.4, and 0.5) GPa.}
    \label{fig:fig_pt}
\end{figure*}

%\textit{Background.}---
Flexural waves propagating within an isotropic homogeneous TEP obey the Kirchhoff-Love biharmonic equation \cite{timoshenko1959theory}, in terms of the vertical displacement $W$, in the frequency-domain regime, i.e., by assuming an $e^{-i\omega t}$ time dependence (See Supplementary Material (SM) \cite{SM} for the general equation in heterogeneous TEPs) $\Delta^2W-\beta^4W=0$, where $\Delta$ is the Laplacian operator \cite{biha}. Moreover, the derivation of the transfer and scattering matrices of this fourth order system are detailed in SM \cite{SM}. This equips us with the necessary mathematical arsenal to fully characterize such layered elastic plate systems in terms of transmission and reflection. In addition to the usual propagating flexural waves, i.e., $e^{i\beta x}$ and $e^{-i\beta x}$, there exist evanescent (inhomogeneous) flexural waves, differentiating the TEP from its acoustic counterpart, in which only the propagating waves are considered. In the free propagation domain, only the propagating component survives as shown in Eq.~(8) in Ref. \cite{SM}, the evanescent wave is proportional to $e^{\beta_L x}$ on the left propagating side (negative $x$) and to $e^{-\beta_R x}$ on the right side (positive $x$). Since these evanescent waves decay exponentially as they travel away from their corresponding interfaces, they do not contribute to the transmission and reflection coefficients, which are measured in the far-field. This is similar to the calculation of the radar scattering cross-section, considered for example in \cite{farhat2014platonic}.
\begin{comment}In particular, at this stage, one remark would be in order, with regards to the evanescent flexural waves. As can be seen from Eq.~(8) in the SM \cite{SM}, the evanescent wave is proportional to $e^{\beta_L x}$ on the left propagating side (negative $x$) and to $e^{-\beta_R x}$ on the right side (positive $x$). In both cases, these waves tend exponentially to zero, as we get further away from the corresponding interface. In order to calculate the reflection and transmission, we need to measure the energy (reflected or transmitted) very far away from the scattering structure, similar to the calculation of the radar scattering cross-section, considered for example in \cite{farhat2014platonic}. This simply means that the coefficients $r_2$ and $t_2$ appearing in the SM [Eqs.~(20)-(24)] \cite{SM} and measured at the interfaces of the structure have no physical meaning far away from the interface, as they have no contribution to the scattered (reflected and/or transmitted) energy.
\end{comment}
However, in order to fully characterize the transmission and reflection of flexural waves, one has to take into account the contribution of all waves at the inner interfaces (shown in Fig. 1 in the SM \cite{SM}). What is more intriguing is that evanescent waves establish propagating components, in the presence of gain and loss. This behavior is contrary to the case of elastic plates without loss and /or gain, where the evanescent waves are confined to the interfaces. \begin{comment} In the present case, these waves travel between the layers and contribute to the scattering from the multi-layered structure. However, in the far-field, one takes into account only the propagating flexural waves.\end{comment}

%\label{sec:section3}

%%%%%%%%%%%%%%%%%%%%%%%%%%%%%%%%%%%%%%%%%%%%%%%%%%%%%%%%%%%%%%%%%%%%%%%%%%%%%%%%%%%%%%%%%

%\subsection{Proof-of-concept}
The structure which we consider (schematized in Fig.~\ref{fig:fig_pt}(a)) consists of three elastic layers denoted as G, L, and P, which stand for gain, loss, and passive, respectively. The possible realization of gain and loss in such elastic structures has been proposed in Refs. \cite{hou2018tunable,hou2018p}. A shunted piezoelectric TEP \cite{vasseur2007waveguiding,hladky1993finite} may lead to an effective Young modulus (of flexural rigidity) with a positive (loss) or negative (gain) imaginary part, depending on the use of an inductor and a positive (negative) resistor. We thus assume that the gain and/or loss can be tuned in a reasonable range. The geometry of the structure is given in Ref. \cite{plate}. The eigenvalues and the reflection and transmission spectra of this structure are computed using the $S$-matrix (See SM \cite{SM}) when a plane flexural waves is impinging from the left and/or the right. The results are depicted in Fig.~\ref{fig:fig_pt}(b)-(c), in the frequency range 10-40 Hz. Since $\mathcal{PT}$-symmetric wave systems are reciprocal, the transmittance is the same for wave incident from both directions. However, for this specific $\mathcal{PT}$-symmetric scenario, the reflectance is drastically different for the right ($R_R$) and left ($R_L$) incidences as shown in the SM \cite{SM} (Fig.~2). The two are related to the transmittance through $r_Lr_R^{*}+tt^{*}=1$. The eigenvalues of the scattering matrix ($s_{\pm}$) are obtained as function of the S-parameters ($t, r_R,$ and $r_L$), i.e. $s_\pm=t\pm\sqrt{r_Lr_R}=t(1\pm i\sqrt{(1/|t|^2-1)})$ (See SM for details \cite{SM}).
\begin{figure*}[t!]
    \centering
    \includegraphics[width=1.65\columnwidth]{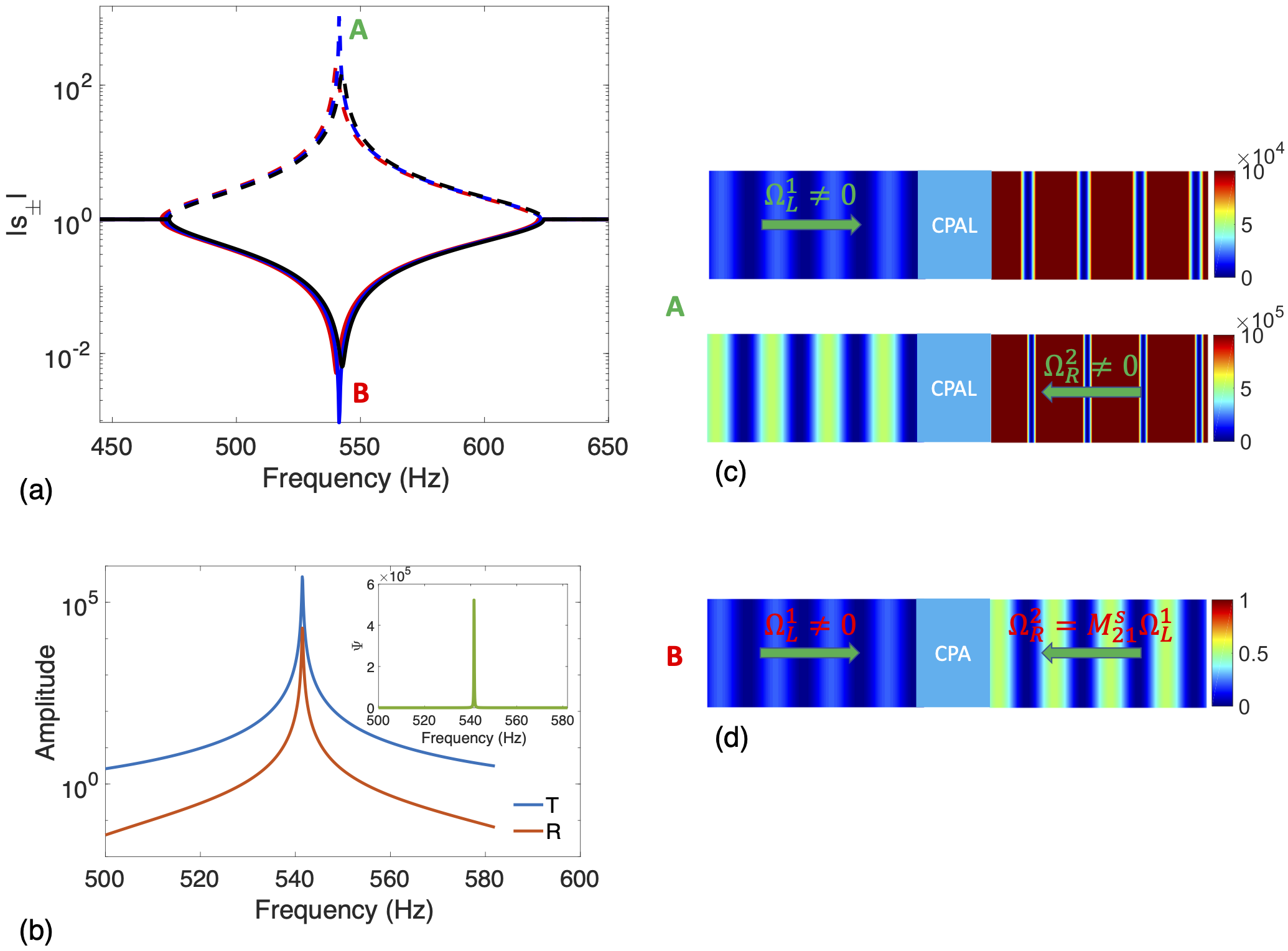}
    \caption{(a) Amplitude of the eigenvalues in the frequency domain where CPAL takes place. Points A and B indicate lasing and perfect absorption operation, respectively. (b) Transmittance and reflectance from the CPAL structure. The inset plots the output coefficient $\Psi$. (c) Snapshots of flexural energy for the $\mathcal{PT}$-symmetric CPA flexural laser in (a) at operating frequency 541 Hz indicated by A in Fig.~\ref{fig:fig_cpal}(a), operated in the lasing mode when the incident wave impinges from the left (top) and the right (bottom). (c) Same as in (a) but for the CPA mode (indicated by B in Fig.~\ref{fig:fig_cpal}(a)) at the same frequency but with both left and right incidence related through $\Omega_R^2=M_{21}^s\Omega_L^1$. The colorbar in (c) and (d) is normalized by the amplitude of the incident flexural wave.}
    \label{fig:fig_cpal}
\end{figure*}
Figures~\ref{fig:fig_pt}(b)-(c) depict the absolute value of the two (propagating) eigenvalues versus the spectral range considered earlier for different values of $\Im(E)$ (highlighted curves show the case of $\Im(E)=0.25$). For frequencies lower than 31 Hz (of the highlighted curves), one can observe non unit-modular eigenvalues. This stems from the fact that $|t|^2>1$, and the flexural system is thus in the so-called broken phase. On the contrary, for frequencies higher than 31 Hz, $s_\pm$ have both unit-module and are non-degenerate, implying that the system is in the symmetric phase. Around this critical frequency, a sudden phase change occurs, whence the $\mathcal{PT}$-symmetric structure flips from a broken-$\mathcal{PT}$ to a $\mathcal{PT}$-symmetric domain: an EP takes place. This EP means a sudden change in the output of the elastic system due to spontaneous breakdown of the $\mathcal{PT}$-symmetry. 
For a small value of $\Im{(E)}$, the EP frequency is around 10 Hz, while for $\Im{(E)}=0.5$ GPa, a blueshift close to 36 Hz can be observed. This behavior is confirmed by observing the phase of $r_L$, that undergoes an abrupt jump of $\pi$-radians, around the same frequencies \cite{SM}, validating the possibility of tuning the EP location by varying the imaginary part of the Young's modulus of gain/loss layers. Such a large tunability of the EP with the amount of (equal) loss (and/or gain) in Young's modulus is somehow specific to flexural waves, as in acoustics, for example, the location of the EP changes only slightly with $\Im{(\rho)}$ (less than $10 \%$ change in the EP frequency compared to $400 \%$ for the flexural case for an equivalent change in the relative imaginary part; also for acoustics the frequency is redshifted with increasing imaginary part, while here it is blueshifted). More detailed analysis of the peculiarity of flexural $\mathcal{PT}$-symmetric systems is given in SM \cite{SM} and showcases more degrees of freedom to tune and thus control the location and even shape of the EP zone, in comparison to other wave systems, essentially due to its parabolic dispersion relation and the coupling between propagating and evanescent waves at the interfaces between the gain/loss layers.

%\textit{CPAL for elasticity.}---
Inspired by this behavior of EP for flexural waves, we consider now the possibility of CPAL effect. In fact, it is well known that in optics, $\mathcal{PT}$-symmetric systems can operate as coherent perfect absorbers by totally absorbing the incoming energy (from impinging waves) and as lasing oscillators by emitting coherently outgoing waves \cite{longhi2010pt,chong2011p,sakhdari2018low}. These two phenomena can be characterized by the overall output flexural coefficient $\Psi$
\begin{equation}
\Psi=\frac{|\Omega_L^2|^2+|\Omega_R^1|^2}{|\Omega_L^1|^2+|\Omega_R^2|^2}\, ,
\label{eq:out_coef}    
\end{equation}
which accounts for the ratio of total outgoing intensity (energy that exits the system) to that of the incoming waves (energy that impinges onto the system). As absorbing and even perfectly-absorbing structures are already known in elasticity and acoustics, we focus here on the effect of lasing for these waves, and that represents the main novelty of this Letter. In the previous results, at best the eigenvalues are offset by 15$\%$. However, for structures consisting of a few unit-cells shown in Fig. 6 of SM \cite{SM}, we can see that more offset may be observed, but the values are still too low to be considered as lasing. To obtain efficient lasing, we maintain the same structure as before (i.e., the three layers shown in Fig.~\ref{fig:fig_pt}) and we apply higher loss/gain parameter, as well as shift to higher (blueshifted) frequencies. The result is plotted in Fig.~\ref{fig:fig_cpal}(a), that depicts the eigenvalues ($|s_\pm|$) for frequencies in the range 450 Hz-650 Hz, corresponding to wavelengths between 18.8 cm and 22.6 cm (to be compared to the width of these layers (in the $x$-direction) assumed to be identical and set as 20 cm). The plot is given for three values of $\Im(E_{g,l})$ (1.49, 1.5, and 1.51 GPa) and $|s_+|$ can reach lasing regime, with $|s_+|>10^3$, at frequency around 541 Hz (denoted A). The complementary eigenvalue $|s_-|\approx0$, which is reminiscent of CPA regime (as in fact we have $s_-=1/s_+^*$). Thus, at the same frequency, one has the lasing regime ($s_+$) where the outgoing energy is hugely amplified and CPA regime ($s_-$) where the outgoing energy is cancelled (i.e., all impinging signal is absorbed by the plate system).

To get a clearer picture, we consider the behavior of the CPAL system operating at the frequency of point A (highlighted in Fig.~\ref{fig:fig_cpal}(a)) under an incident flexural wave of unit-amplitude $|\Omega_{L,R}^{1,2}|=1$. The transmitted and reflected flexural energy $T$ and $R$ are depicted in Fig.~\ref{fig:fig_cpal}(b). At the CPAL point, both $T$ and $R$ reach extremely high values. The coefficient $\Psi$, shown in the inset of Fig.~\ref{fig:fig_cpal}(b), gives a better picture of the lasing efficiency of the device. At frequency 541.4 Hz the spike reaching $10^6$ is a clear evidence of the lasing effect. To further demonstrate this effect, in Fig.~\ref{fig:fig_cpal}(c) we plot the flexural energy in the vicinity of the CPAL device where a normally incident flexural wave of unit-amplitude is impinging from the left (top panel) and the right (bottom panel). In both cases, the outgoing waves (reflected and transmitted) are significantly amplified (in the range of $10^5$ to $10^6$ depending on the incidence region). However, for one scenario (left incidence) the transmittance is higher than the reflectance. For the other scenario (right incidence) it is the reflectance which is higher as can be seen from the plots.

These results demonstrate the potential of using simple and compact (60 cm width and 2 cm thickness $\delta$) flexural systems to achieve the equivalence of a flexural laser that we might call a FLASER. Consider a flexural wave with very small vertical displacement, of amplitude $|W|\approx10\,\mu\textrm{m}$ (i.e., $|W|\ll\delta$) incident on the CPAL. Although this signal is small, it will be amplified by the CPAL flexural device, and the output displacement will be in the range of 1 cm, i.e., $|W|\approx\delta$. Now to relate this effect to the transfer matrix (See SM \cite{SM}) it is straightforward to see that lasing may occur, when we have finite (propagating) outgoing signals $\Omega_L^2$ and $\Omega_R^1$ for very small incoming signals. This may occur for $M^s_{22}=0$ (and $M^s_{12}=\Omega_R^1/\Omega_L^2$) if we ignore the evanescent fields, by inspection of Eqs. (21)-(24) in the SM \cite{SM}. However, this is generally not possible, as the evanescent fields $\Omega_L^4$ and $\Omega_R^3$ cannot be assumed to be zero at the boundary of the system. Hence, the reduced system (in terms of reflection coefficients) to be satisfied in the general case, i.e., by asking that the incident signal has unit-amplitude ($\Omega_L^1=1$) and that the outgoing signals diverge, can b expressed as
\begin{equation}
   \begin{pmatrix}
  M_{22}^s & M_{24}^s\\
  M_{42}^s & M_{44}^s  
 \end{pmatrix}\begin{pmatrix}
  \Omega_L^2 \\
  \Omega_L^4
 \end{pmatrix}=-\begin{pmatrix}
  M_{24}^s \\
  M_{41}^s
 \end{pmatrix}\, .
\label{eq:lasing}
\end{equation}

\begin{figure}[b!]
    \centering
    \includegraphics[width=1\columnwidth]{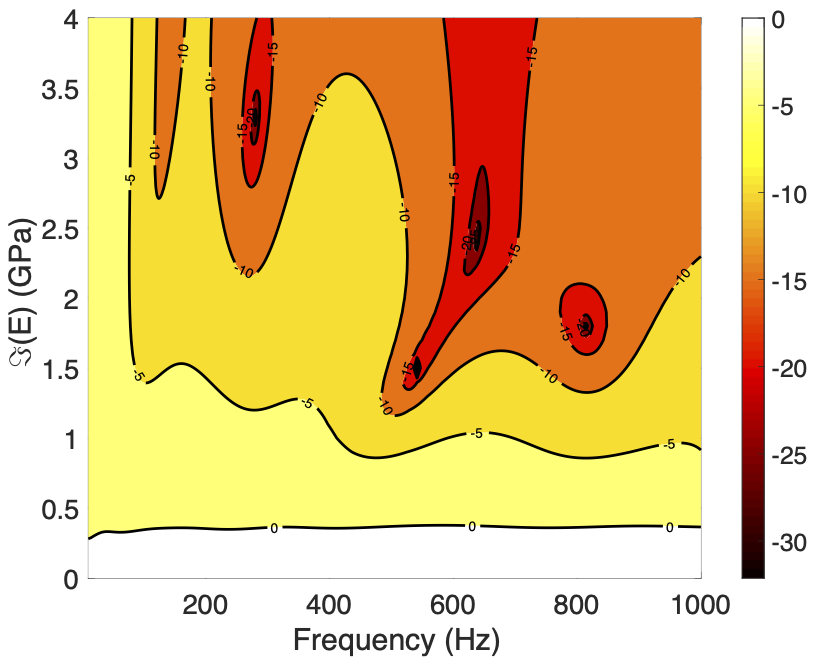}
    \caption{Contourplot of the variation of the parameter $\Delta M^s=M_{22}^sM_{44}^s-M_{42}^sM_{24}^s$ versus the frequency and landscape of the imaginary part of the Young's modulus (in GPa), in logarithmic scale. The dark regions correspond to the lasing regime (i.e., $\Delta M^s\approx0$).}
    \label{fig:fig_cpal_2d}
\end{figure}

If the incidence is taken to be zero and if one imposes finite scattering signals, as occurs in lasing, one must thus ensure that the determinant of the matrix in the LHS of Eq.~(\ref{eq:lasing}) is zero, which yields $M_{22}^sM_{44}^s-M_{42}^sM_{24}^s=\Delta M^s=0$, which is markedly different from the simple condition $M_{22}^s=0$ for acoustic or optical systems, for example. For the transmitted signals, it is easy to obtain their expression, by taking the first and third line in Eq. (22) of the SM \cite{SM}. This gives for example $\Omega_R^1=M_{12}^s\Omega_L^2+M_{14}^s\Omega_L^4$. This complexity stems thus from the interplay between propagating and evanescent waves that cannot be ignored for flexural systems, as clearly demonstrated by the lasing equation that mixes amplitudes of both kinds of waves. The variation of the parameter $\Delta M^s$ responsible for infinite outgoing amplitudes (i.e., lasing) is depicted in Fig.~\ref{fig:fig_cpal_2d} versus the frequency of the flexural wave and the imaginary part of the young modulus (in GPa). The dark regions correspond to the lasing regime (i.e., $\Delta M^s\approx0$). One can clearly see that if lasing is defined when the logarithmic amplitude of the lasing parameter is below -20 dB, a lasing threshold in terms of $\Im(E)$ can be defined for each frequency, below which no lasing can occur. From this landscape of the imaginary part of Young's modulus it can also be seen that the lowest threshold occurs for a frequency of 541 Hz, as was discussed previously.  

A similar reasoning can be made for the CPA effect. In fact, for perfect absorption to occur, one must cancel the outgoing waves for finite incoming waves. The same analysis as before shows that for CPA to take place, one must ensure that $M_{11}=0$ and $M^s_{21}=\Omega_R^2/\Omega_L^1$, overlooking again evanescent waves. However, in the general case, if also an evanescent wave is allowed to be incident, one must ensure that $M^s_{11}\Omega_L^1+M^s_{13}\Omega_L^3=0$ and $M^s_{21}\Omega_L^1+M^s_{23}\Omega_L^3=\Omega_R^2$. Yet, in the CPA case, $\Omega_L^3=0$ as it corresponds to an exponentially growing field when $x\rightarrow-\infty$. Thus for the CPA operation, evanescent waves are not directly present in the condition on the amplitudes. However, their indirect effect in the $M$-matrix is still present. Therefore, one needs to launch two waves incoming from opposite directions, and by changing their (complex) amplitude ratio, one will be able to selectively excite the lasing or CPA mode, as can be seen in Fig.~\ref{fig:fig_cpal}(d). If their amplitude ratio $M^s_{21}=\Omega_R^2/\Omega_L^1$, one will achieve the CPA operation mode. In the rest of the cases, i.e., $M^s_{21}\neq\Omega_R^2/\Omega_L^1$ one will have the FLASER mode. This is, however, a very narrowband effect for both CPA and lasing. 

%%%%%%%%%%%%%%%%%%%%%%%%%%%%%%%%%%%%%%%%%%%%%%%%%%%%%%%%%%%%%%%%%%%%%%%%%%%%%%%%%%%%%%%%%%%%%%%%%%%
%\label{sec:section4}
%\textit{Discussion and summary.}---
To summarize, an effect reminiscent of lasing is discovered in this Letter by making use of $\mathcal{PT}$-symmetry in a specific frequency and gain/loss regimes. This mechanism takes roots in the coherent perfect absorber laser effect. This CPAL device can be used as an ultra-sensitive flexural sensor to detect sub-micrometer displacements or as a perfect absorber of flexural energy. However, in stark contrast to Maxwell's equations, the flexural wave equation assumes displacements that are small in comparison with the thickness of the plate, so our device could be used as a source, since by applying very small displacements, these will be amplified and coherently transmitted  to  the  outside. On the detection side, if some strong signal/noise impinges on the device, it can be hugely amplified and can lead to its dislocation. One way to avoid this unwanted effect is to use a filter that cancels out incoming signals above a certain threshold. Our work can thus pave the way to several interesting applications in surface science and civil engineering, e.g. in precision-displacement sensing, vibration control of mechanical systems, and seismic energy harvesting. 

%%%%%%%%%%%%%%%%%%%%%%%%%%%%%%%%%%%%%%%%%%%%%%%%%%%%%%%%%%%%%%%%%%%%%%%%%%%%%%%%%%%%%%%%%%%%%%%%%%%
The research reported in this manuscript was supported by King Abdullah University of Science and Technology, Baseline Research Fund BAS/1/1626-01-01.

%This work is supported by King Abdullah University of Science and Technology (KAUST) under Baseline Research Fund BAS/1/1626-01-01. 

%%%%%%%%%%%%%%%%%%%%%%%%%%%%%%%%%%%%%%%%%%%%%%%%%%%%%%%%%%%%%%%%%%%%%%%%%%%%%%%%%%%%%%%%%%%%%%%%%%%

\nocite{*}

\bibliography{References}% Produces the bibliography via BibTeX.

\providecommand{\noopsort}[1]{}\providecommand{\singleletter}[1]{#1}%
\begin{thebibliography}{43}%
\makeatletter
\providecommand \@ifxundefined [1]{%
 \@ifx{#1\undefined}
}%
\providecommand \@ifnum [1]{%
 \ifnum #1\expandafter \@firstoftwo
 \else \expandafter \@secondoftwo
 \fi
}%
\providecommand \@ifx [1]{%
 \ifx #1\expandafter \@firstoftwo
 \else \expandafter \@secondoftwo
 \fi
}%
\providecommand \natexlab [1]{#1}%
\providecommand \enquote  [1]{``#1''}%
\providecommand \bibnamefont  [1]{#1}%
\providecommand \bibfnamefont [1]{#1}%
\providecommand \citenamefont [1]{#1}%
\providecommand \href@noop [0]{\@secondoftwo}%
\providecommand \href [0]{\begingroup \@sanitize@url \@href}%
\providecommand \@href[1]{\@@startlink{#1}\@@href}%
\providecommand \@@href[1]{\endgroup#1\@@endlink}%
\providecommand \@sanitize@url [0]{\catcode `\\12\catcode `\$12\catcode
  `\&12\catcode `\#12\catcode `\^12\catcode `\_12\catcode `\%12\relax}%
\providecommand \@@startlink[1]{}%
\providecommand \@@endlink[0]{}%
\providecommand \url  [0]{\begingroup\@sanitize@url \@url }%
\providecommand \@url [1]{\endgroup\@href {#1}{\urlprefix }}%
\providecommand \urlprefix  [0]{URL }%
\providecommand \Eprint [0]{\href }%
\providecommand \doibase [0]{https://doi.org/}%
\providecommand \selectlanguage [0]{\@gobble}%
\providecommand \bibinfo  [0]{\@secondoftwo}%
\providecommand \bibfield  [0]{\@secondoftwo}%
\providecommand \translation [1]{[#1]}%
\providecommand \BibitemOpen [0]{}%
\providecommand \bibitemStop [0]{}%
\providecommand \bibitemNoStop [0]{.\EOS\space}%
\providecommand \EOS [0]{\spacefactor3000\relax}%
\providecommand \BibitemShut  [1]{\csname bibitem#1\endcsname}%
\let\auto@bib@innerbib\@empty
%</preamble>
\bibitem [{\citenamefont {Pendry}\ \emph {et~al.}(1999)\citenamefont {Pendry},
  \citenamefont {Holden}, \citenamefont {Robbins},\ and\ \citenamefont
  {Stewart}}]{pendry1999magnetism}%
  \BibitemOpen
  \bibfield  {author} {\bibinfo {author} {\bibfnamefont {J.~B.}\ \bibnamefont
  {Pendry}}, \bibinfo {author} {\bibfnamefont {A.~J.}\ \bibnamefont {Holden}},
  \bibinfo {author} {\bibfnamefont {D.~J.}\ \bibnamefont {Robbins}},\ and\
  \bibinfo {author} {\bibfnamefont {W.}~\bibnamefont {Stewart}},\ }\bibfield
  {title} {\bibinfo {title} {Magnetism from conductors and enhanced nonlinear
  phenomena},\ }\href@noop {} {\bibfield  {journal} {\bibinfo  {journal} {IEEE
  transactions on microwave theory and techniques}\ }\textbf {\bibinfo {volume}
  {47}},\ \bibinfo {pages} {2075} (\bibinfo {year} {1999})}\BibitemShut
  {NoStop}%
\bibitem [{\citenamefont {Pendry}(2000)}]{pendry2000negative}%
  \BibitemOpen
  \bibfield  {author} {\bibinfo {author} {\bibfnamefont {J.~B.}\ \bibnamefont
  {Pendry}},\ }\bibfield  {title} {\bibinfo {title} {Negative refraction makes
  a perfect lens},\ }\href@noop {} {\bibfield  {journal} {\bibinfo  {journal}
  {Physical Review Letters}\ }\textbf {\bibinfo {volume} {85}},\ \bibinfo
  {pages} {3966} (\bibinfo {year} {2000})}\BibitemShut {NoStop}%
\bibitem [{\citenamefont {Smith}\ \emph {et~al.}(2004)\citenamefont {Smith},
  \citenamefont {Pendry},\ and\ \citenamefont
  {Wiltshire}}]{smith2004metamaterials}%
  \BibitemOpen
  \bibfield  {author} {\bibinfo {author} {\bibfnamefont {D.~R.}\ \bibnamefont
  {Smith}}, \bibinfo {author} {\bibfnamefont {J.~B.}\ \bibnamefont {Pendry}},\
  and\ \bibinfo {author} {\bibfnamefont {M.~C.}\ \bibnamefont {Wiltshire}},\
  }\bibfield  {title} {\bibinfo {title} {Metamaterials and negative refractive
  index},\ }\href@noop {} {\bibfield  {journal} {\bibinfo  {journal} {Science}\
  }\textbf {\bibinfo {volume} {305}},\ \bibinfo {pages} {788} (\bibinfo {year}
  {2004})}\BibitemShut {NoStop}%
\bibitem [{\citenamefont {Al{\`u}}\ and\ \citenamefont
  {Engheta}(2005)}]{alu2005achieving}%
  \BibitemOpen
  \bibfield  {author} {\bibinfo {author} {\bibfnamefont {A.}~\bibnamefont
  {Al{\`u}}}\ and\ \bibinfo {author} {\bibfnamefont {N.}~\bibnamefont
  {Engheta}},\ }\bibfield  {title} {\bibinfo {title} {Achieving transparency
  with plasmonic and metamaterial coatings},\ }\href@noop {} {\bibfield
  {journal} {\bibinfo  {journal} {Physical Review E}\ }\textbf {\bibinfo
  {volume} {72}},\ \bibinfo {pages} {016623} (\bibinfo {year}
  {2005})}\BibitemShut {NoStop}%
\bibitem [{\citenamefont {Chen}\ \emph {et~al.}(2012)\citenamefont {Chen},
  \citenamefont {Soric},\ and\ \citenamefont {Alu}}]{chen2012invisibility}%
  \BibitemOpen
  \bibfield  {author} {\bibinfo {author} {\bibfnamefont {P.-Y.}\ \bibnamefont
  {Chen}}, \bibinfo {author} {\bibfnamefont {J.}~\bibnamefont {Soric}},\ and\
  \bibinfo {author} {\bibfnamefont {A.}~\bibnamefont {Alu}},\ }\bibfield
  {title} {\bibinfo {title} {Invisibility and cloaking based on scattering
  cancellation},\ }\href@noop {} {\bibfield  {journal} {\bibinfo  {journal}
  {Advanced Materials}\ }\textbf {\bibinfo {volume} {24}},\ \bibinfo {pages}
  {OP281} (\bibinfo {year} {2012})}\BibitemShut {NoStop}%
\bibitem [{\citenamefont {Movchan}\ \emph {et~al.}(2007)\citenamefont
  {Movchan}, \citenamefont {Movchan},\ and\ \citenamefont
  {McPhedran}}]{movchan2007bloch}%
  \BibitemOpen
  \bibfield  {author} {\bibinfo {author} {\bibfnamefont {A.}~\bibnamefont
  {Movchan}}, \bibinfo {author} {\bibfnamefont {N.}~\bibnamefont {Movchan}},\
  and\ \bibinfo {author} {\bibfnamefont {R.}~\bibnamefont {McPhedran}},\
  }\bibfield  {title} {\bibinfo {title} {Bloch--floquet bending waves in
  perforated thin plates},\ }\href@noop {} {\bibfield  {journal} {\bibinfo
  {journal} {Proceedings of the Royal Society A: Mathematical, Physical and
  Engineering Sciences}\ }\textbf {\bibinfo {volume} {463}},\ \bibinfo {pages}
  {2505} (\bibinfo {year} {2007})}\BibitemShut {NoStop}%
\bibitem [{\citenamefont {Wu}\ \emph {et~al.}(2011)\citenamefont {Wu},
  \citenamefont {Lai},\ and\ \citenamefont {Zhang}}]{wu2011elastic}%
  \BibitemOpen
  \bibfield  {author} {\bibinfo {author} {\bibfnamefont {Y.}~\bibnamefont
  {Wu}}, \bibinfo {author} {\bibfnamefont {Y.}~\bibnamefont {Lai}},\ and\
  \bibinfo {author} {\bibfnamefont {Z.-Q.}\ \bibnamefont {Zhang}},\ }\bibfield
  {title} {\bibinfo {title} {Elastic metamaterials with simultaneously negative
  effective shear modulus and mass density},\ }\href@noop {} {\bibfield
  {journal} {\bibinfo  {journal} {Physical Review Letters}\ }\textbf {\bibinfo
  {volume} {107}},\ \bibinfo {pages} {105506} (\bibinfo {year}
  {2011})}\BibitemShut {NoStop}%
\bibitem [{\citenamefont {Timoshenko}\ and\ \citenamefont
  {Woinowsky-Krieger}(1959)}]{timoshenko1959theory}%
  \BibitemOpen
  \bibfield  {author} {\bibinfo {author} {\bibfnamefont {S.~P.}\ \bibnamefont
  {Timoshenko}}\ and\ \bibinfo {author} {\bibfnamefont {S.}~\bibnamefont
  {Woinowsky-Krieger}},\ }\href@noop {} {\emph {\bibinfo {title} {Theory of
  plates and shells}}}\ (\bibinfo  {publisher} {McGraw-hill},\ \bibinfo {year}
  {1959})\BibitemShut {NoStop}%
\bibitem [{\citenamefont {Norris}\ and\ \citenamefont
  {Vemula}(1995)}]{norris1995scattering}%
  \BibitemOpen
  \bibfield  {author} {\bibinfo {author} {\bibfnamefont {A.}~\bibnamefont
  {Norris}}\ and\ \bibinfo {author} {\bibfnamefont {C.}~\bibnamefont
  {Vemula}},\ }\bibfield  {title} {\bibinfo {title} {Scattering of flexural
  waves on thin plates},\ }\href@noop {} {\bibfield  {journal} {\bibinfo
  {journal} {Journal of sound and vibration}\ }\textbf {\bibinfo {volume}
  {181}},\ \bibinfo {pages} {115} (\bibinfo {year} {1995})}\BibitemShut
  {NoStop}%
\bibitem [{\citenamefont {Farhat}\ \emph {et~al.}(2009)\citenamefont {Farhat},
  \citenamefont {Guenneau},\ and\ \citenamefont
  {Enoch}}]{farhat2009ultrabroadband}%
  \BibitemOpen
  \bibfield  {author} {\bibinfo {author} {\bibfnamefont {M.}~\bibnamefont
  {Farhat}}, \bibinfo {author} {\bibfnamefont {S.}~\bibnamefont {Guenneau}},\
  and\ \bibinfo {author} {\bibfnamefont {S.}~\bibnamefont {Enoch}},\ }\bibfield
   {title} {\bibinfo {title} {Ultrabroadband elastic cloaking in thin plates},\
  }\href@noop {} {\bibfield  {journal} {\bibinfo  {journal} {Physical Review
  Letters}\ }\textbf {\bibinfo {volume} {103}},\ \bibinfo {pages} {024301}
  (\bibinfo {year} {2009})}\BibitemShut {NoStop}%
\bibitem [{\citenamefont {Zhu}\ \emph {et~al.}(2018)\citenamefont {Zhu},
  \citenamefont {Liu}, \citenamefont {Liang}, \citenamefont {Chen},\ and\
  \citenamefont {Li}}]{zhu2018elastic}%
  \BibitemOpen
  \bibfield  {author} {\bibinfo {author} {\bibfnamefont {J.}~\bibnamefont
  {Zhu}}, \bibinfo {author} {\bibfnamefont {Y.}~\bibnamefont {Liu}}, \bibinfo
  {author} {\bibfnamefont {Z.}~\bibnamefont {Liang}}, \bibinfo {author}
  {\bibfnamefont {T.}~\bibnamefont {Chen}},\ and\ \bibinfo {author}
  {\bibfnamefont {J.}~\bibnamefont {Li}},\ }\bibfield  {title} {\bibinfo
  {title} {Elastic waves in curved space: mimicking a wormhole},\ }\href@noop
  {} {\bibfield  {journal} {\bibinfo  {journal} {Physical Review Letters}\
  }\textbf {\bibinfo {volume} {121}},\ \bibinfo {pages} {234301} (\bibinfo
  {year} {2018})}\BibitemShut {NoStop}%
\bibitem [{\citenamefont {Dubois}\ \emph {et~al.}(2013)\citenamefont {Dubois},
  \citenamefont {Farhat}, \citenamefont {Bossy}, \citenamefont {Enoch},
  \citenamefont {Guenneau},\ and\ \citenamefont {Sebbah}}]{dubois2014flat}%
  \BibitemOpen
  \bibfield  {author} {\bibinfo {author} {\bibfnamefont {M.}~\bibnamefont
  {Dubois}}, \bibinfo {author} {\bibfnamefont {M.}~\bibnamefont {Farhat}},
  \bibinfo {author} {\bibfnamefont {E.}~\bibnamefont {Bossy}}, \bibinfo
  {author} {\bibfnamefont {S.}~\bibnamefont {Enoch}}, \bibinfo {author}
  {\bibfnamefont {S.}~\bibnamefont {Guenneau}},\ and\ \bibinfo {author}
  {\bibfnamefont {P.}~\bibnamefont {Sebbah}},\ }\bibfield  {title} {\bibinfo
  {title} {Flat lens for pulse focusing of elastic waves in thin plates},\
  }\href {https://doi.org/10.1063/1.4818716} {\bibfield  {journal} {\bibinfo
  {journal} {Applied Physics Letters}\ }\textbf {\bibinfo {volume} {103}},\
  \bibinfo {pages} {071915} (\bibinfo {year} {2013})}\BibitemShut {NoStop}%
\bibitem [{\citenamefont {Farhat}\ \emph {et~al.}(2017)\citenamefont {Farhat},
  \citenamefont {Chen}, \citenamefont {Guenneau}, \citenamefont {Salama},\ and\
  \citenamefont {Ba{\u{g}}c{\i}}}]{farhat2017localized}%
  \BibitemOpen
  \bibfield  {author} {\bibinfo {author} {\bibfnamefont {M.}~\bibnamefont
  {Farhat}}, \bibinfo {author} {\bibfnamefont {P.-Y.}\ \bibnamefont {Chen}},
  \bibinfo {author} {\bibfnamefont {S.}~\bibnamefont {Guenneau}}, \bibinfo
  {author} {\bibfnamefont {K.~N.}\ \bibnamefont {Salama}},\ and\ \bibinfo
  {author} {\bibfnamefont {H.}~\bibnamefont {Ba{\u{g}}c{\i}}},\ }\bibfield
  {title} {\bibinfo {title} {Localized surface plate modes via flexural mie
  resonances},\ }\href@noop {} {\bibfield  {journal} {\bibinfo  {journal}
  {Physical Review B}\ }\textbf {\bibinfo {volume} {95}},\ \bibinfo {pages}
  {174201} (\bibinfo {year} {2017})}\BibitemShut {NoStop}%
\bibitem [{\citenamefont {Farhat}\ \emph {et~al.}(2014)\citenamefont {Farhat},
  \citenamefont {Chen}, \citenamefont {Ba{\u{g}}c{\i}}, \citenamefont {Enoch},
  \citenamefont {Guenneau},\ and\ \citenamefont {Alu}}]{farhat2014platonic}%
  \BibitemOpen
  \bibfield  {author} {\bibinfo {author} {\bibfnamefont {M.}~\bibnamefont
  {Farhat}}, \bibinfo {author} {\bibfnamefont {P.-Y.}\ \bibnamefont {Chen}},
  \bibinfo {author} {\bibfnamefont {H.}~\bibnamefont {Ba{\u{g}}c{\i}}},
  \bibinfo {author} {\bibfnamefont {S.}~\bibnamefont {Enoch}}, \bibinfo
  {author} {\bibfnamefont {S.}~\bibnamefont {Guenneau}},\ and\ \bibinfo
  {author} {\bibfnamefont {A.}~\bibnamefont {Alu}},\ }\bibfield  {title}
  {\bibinfo {title} {Platonic scattering cancellation for bending waves in a
  thin plate},\ }\href@noop {} {\bibfield  {journal} {\bibinfo  {journal}
  {Scientific reports}\ }\textbf {\bibinfo {volume} {4}},\ \bibinfo {pages}
  {4644} (\bibinfo {year} {2014})}\BibitemShut {NoStop}%
\bibitem [{\citenamefont {McPhedran}\ \emph {et~al.}(2015)\citenamefont
  {McPhedran}, \citenamefont {Movchan}, \citenamefont {Movchan}, \citenamefont
  {Brun},\ and\ \citenamefont {Smith}}]{mcphedran2015parabolic}%
  \BibitemOpen
  \bibfield  {author} {\bibinfo {author} {\bibfnamefont {R.}~\bibnamefont
  {McPhedran}}, \bibinfo {author} {\bibfnamefont {A.}~\bibnamefont {Movchan}},
  \bibinfo {author} {\bibfnamefont {N.}~\bibnamefont {Movchan}}, \bibinfo
  {author} {\bibfnamefont {M.}~\bibnamefont {Brun}},\ and\ \bibinfo {author}
  {\bibfnamefont {M.}~\bibnamefont {Smith}},\ }\bibfield  {title} {\bibinfo
  {title} {Parabolic trapped modes and steered dirac cones in platonic
  crystals},\ }\href@noop {} {\bibfield  {journal} {\bibinfo  {journal}
  {Proceedings of the Royal Society A: Mathematical, Physical and Engineering
  Sciences}\ }\textbf {\bibinfo {volume} {471}},\ \bibinfo {pages} {20140746}
  (\bibinfo {year} {2015})}\BibitemShut {NoStop}%
\bibitem [{\citenamefont {Bender}\ and\ \citenamefont
  {Boettcher}(1998)}]{bender1998real}%
  \BibitemOpen
  \bibfield  {author} {\bibinfo {author} {\bibfnamefont {C.~M.}\ \bibnamefont
  {Bender}}\ and\ \bibinfo {author} {\bibfnamefont {S.}~\bibnamefont
  {Boettcher}},\ }\bibfield  {title} {\bibinfo {title} {Real spectra in
  non-hermitian hamiltonians having p t symmetry},\ }\href@noop {} {\bibfield
  {journal} {\bibinfo  {journal} {Physical Review Letters}\ }\textbf {\bibinfo
  {volume} {80}},\ \bibinfo {pages} {5243} (\bibinfo {year}
  {1998})}\BibitemShut {NoStop}%
\bibitem [{\citenamefont {Mostafazadeh}(2002)}]{mostafazadeh2002pseudo}%
  \BibitemOpen
  \bibfield  {author} {\bibinfo {author} {\bibfnamefont {A.}~\bibnamefont
  {Mostafazadeh}},\ }\bibfield  {title} {\bibinfo {title} {Pseudo-hermiticity
  versus pt-symmetry. ii. a complete characterization of non-hermitian
  hamiltonians with a real spectrum},\ }\href@noop {} {\bibfield  {journal}
  {\bibinfo  {journal} {Journal of Mathematical Physics}\ }\textbf {\bibinfo
  {volume} {43}},\ \bibinfo {pages} {2814} (\bibinfo {year}
  {2002})}\BibitemShut {NoStop}%
\bibitem [{\citenamefont {Makris}\ \emph {et~al.}(2008)\citenamefont {Makris},
  \citenamefont {El-Ganainy}, \citenamefont {Christodoulides},\ and\
  \citenamefont {Musslimani}}]{makris2008beam}%
  \BibitemOpen
  \bibfield  {author} {\bibinfo {author} {\bibfnamefont {K.~G.}\ \bibnamefont
  {Makris}}, \bibinfo {author} {\bibfnamefont {R.}~\bibnamefont {El-Ganainy}},
  \bibinfo {author} {\bibfnamefont {D.}~\bibnamefont {Christodoulides}},\ and\
  \bibinfo {author} {\bibfnamefont {Z.~H.}\ \bibnamefont {Musslimani}},\
  }\bibfield  {title} {\bibinfo {title} {Beam dynamics in pt symmetric optical
  lattices},\ }\href@noop {} {\bibfield  {journal} {\bibinfo  {journal}
  {Physical Review Letters}\ }\textbf {\bibinfo {volume} {100}},\ \bibinfo
  {pages} {103904} (\bibinfo {year} {2008})}\BibitemShut {NoStop}%
\bibitem [{\citenamefont {R{\"u}ter}\ \emph {et~al.}(2010)\citenamefont
  {R{\"u}ter}, \citenamefont {Makris}, \citenamefont {El-Ganainy},
  \citenamefont {Christodoulides}, \citenamefont {Segev},\ and\ \citenamefont
  {Kip}}]{ruter2010observation}%
  \BibitemOpen
  \bibfield  {author} {\bibinfo {author} {\bibfnamefont {C.~E.}\ \bibnamefont
  {R{\"u}ter}}, \bibinfo {author} {\bibfnamefont {K.~G.}\ \bibnamefont
  {Makris}}, \bibinfo {author} {\bibfnamefont {R.}~\bibnamefont {El-Ganainy}},
  \bibinfo {author} {\bibfnamefont {D.~N.}\ \bibnamefont {Christodoulides}},
  \bibinfo {author} {\bibfnamefont {M.}~\bibnamefont {Segev}},\ and\ \bibinfo
  {author} {\bibfnamefont {D.}~\bibnamefont {Kip}},\ }\bibfield  {title}
  {\bibinfo {title} {Observation of parity--time symmetry in optics},\
  }\href@noop {} {\bibfield  {journal} {\bibinfo  {journal} {Nature Physics}\
  }\textbf {\bibinfo {volume} {6}},\ \bibinfo {pages} {192} (\bibinfo {year}
  {2010})}\BibitemShut {NoStop}%
\bibitem [{\citenamefont {Feng}\ \emph {et~al.}(2017)\citenamefont {Feng},
  \citenamefont {El-Ganainy},\ and\ \citenamefont {Ge}}]{feng2017non}%
  \BibitemOpen
  \bibfield  {author} {\bibinfo {author} {\bibfnamefont {L.}~\bibnamefont
  {Feng}}, \bibinfo {author} {\bibfnamefont {R.}~\bibnamefont {El-Ganainy}},\
  and\ \bibinfo {author} {\bibfnamefont {L.}~\bibnamefont {Ge}},\ }\bibfield
  {title} {\bibinfo {title} {Non-hermitian photonics based on parity--time
  symmetry},\ }\href@noop {} {\bibfield  {journal} {\bibinfo  {journal} {Nature
  Photonics}\ }\textbf {\bibinfo {volume} {11}},\ \bibinfo {pages} {752}
  (\bibinfo {year} {2017})}\BibitemShut {NoStop}%
\bibitem [{\citenamefont {Sakhdari}\ \emph {et~al.}(2017)\citenamefont
  {Sakhdari}, \citenamefont {Farhat},\ and\ \citenamefont
  {Chen}}]{sakhdari2017pt}%
  \BibitemOpen
  \bibfield  {author} {\bibinfo {author} {\bibfnamefont {M.}~\bibnamefont
  {Sakhdari}}, \bibinfo {author} {\bibfnamefont {M.}~\bibnamefont {Farhat}},\
  and\ \bibinfo {author} {\bibfnamefont {P.-Y.}\ \bibnamefont {Chen}},\
  }\bibfield  {title} {\bibinfo {title} {Pt-symmetric metasurfaces: wave
  manipulation and sensing using singular points},\ }\href@noop {} {\bibfield
  {journal} {\bibinfo  {journal} {New Journal of Physics}\ }\textbf {\bibinfo
  {volume} {19}},\ \bibinfo {pages} {065002} (\bibinfo {year}
  {2017})}\BibitemShut {NoStop}%
\bibitem [{\citenamefont {Chen}\ and\ \citenamefont {Jung}(2016)}]{chen2016p}%
  \BibitemOpen
  \bibfield  {author} {\bibinfo {author} {\bibfnamefont {P.-Y.}\ \bibnamefont
  {Chen}}\ and\ \bibinfo {author} {\bibfnamefont {J.}~\bibnamefont {Jung}},\
  }\bibfield  {title} {\bibinfo {title} {Pt symmetry and singularity-enhanced
  sensing based on photoexcited graphene metasurfaces},\ }\href@noop {}
  {\bibfield  {journal} {\bibinfo  {journal} {Physical Review Applied}\
  }\textbf {\bibinfo {volume} {5}},\ \bibinfo {pages} {064018} (\bibinfo {year}
  {2016})}\BibitemShut {NoStop}%
\bibitem [{\citenamefont {Peng}\ \emph {et~al.}(2014)\citenamefont {Peng},
  \citenamefont {{\"O}zdemir}, \citenamefont {Lei}, \citenamefont {Monifi},
  \citenamefont {Gianfreda}, \citenamefont {Long}, \citenamefont {Fan},
  \citenamefont {Nori}, \citenamefont {Bender},\ and\ \citenamefont
  {Yang}}]{peng2014parity}%
  \BibitemOpen
  \bibfield  {author} {\bibinfo {author} {\bibfnamefont {B.}~\bibnamefont
  {Peng}}, \bibinfo {author} {\bibfnamefont {{\c{S}}.~K.}\ \bibnamefont
  {{\"O}zdemir}}, \bibinfo {author} {\bibfnamefont {F.}~\bibnamefont {Lei}},
  \bibinfo {author} {\bibfnamefont {F.}~\bibnamefont {Monifi}}, \bibinfo
  {author} {\bibfnamefont {M.}~\bibnamefont {Gianfreda}}, \bibinfo {author}
  {\bibfnamefont {G.~L.}\ \bibnamefont {Long}}, \bibinfo {author}
  {\bibfnamefont {S.}~\bibnamefont {Fan}}, \bibinfo {author} {\bibfnamefont
  {F.}~\bibnamefont {Nori}}, \bibinfo {author} {\bibfnamefont {C.~M.}\
  \bibnamefont {Bender}},\ and\ \bibinfo {author} {\bibfnamefont
  {L.}~\bibnamefont {Yang}},\ }\bibfield  {title} {\bibinfo {title}
  {Parity--time-symmetric whispering-gallery microcavities},\ }\href@noop {}
  {\bibfield  {journal} {\bibinfo  {journal} {Nature Physics}\ }\textbf
  {\bibinfo {volume} {10}},\ \bibinfo {pages} {394} (\bibinfo {year}
  {2014})}\BibitemShut {NoStop}%
\bibitem [{\citenamefont {Feng}\ \emph {et~al.}(2014)\citenamefont {Feng},
  \citenamefont {Wong}, \citenamefont {Ma}, \citenamefont {Wang},\ and\
  \citenamefont {Zhang}}]{feng2014single}%
  \BibitemOpen
  \bibfield  {author} {\bibinfo {author} {\bibfnamefont {L.}~\bibnamefont
  {Feng}}, \bibinfo {author} {\bibfnamefont {Z.~J.}\ \bibnamefont {Wong}},
  \bibinfo {author} {\bibfnamefont {R.-M.}\ \bibnamefont {Ma}}, \bibinfo
  {author} {\bibfnamefont {Y.}~\bibnamefont {Wang}},\ and\ \bibinfo {author}
  {\bibfnamefont {X.}~\bibnamefont {Zhang}},\ }\bibfield  {title} {\bibinfo
  {title} {Single-mode laser by parity-time symmetry breaking},\ }\href@noop {}
  {\bibfield  {journal} {\bibinfo  {journal} {Science}\ }\textbf {\bibinfo
  {volume} {346}},\ \bibinfo {pages} {972} (\bibinfo {year}
  {2014})}\BibitemShut {NoStop}%
\bibitem [{\citenamefont {Zhu}\ \emph {et~al.}(2014)\citenamefont {Zhu},
  \citenamefont {Ramezani}, \citenamefont {Shi}, \citenamefont {Zhu},\ and\
  \citenamefont {Zhang}}]{zhu2014p}%
  \BibitemOpen
  \bibfield  {author} {\bibinfo {author} {\bibfnamefont {X.}~\bibnamefont
  {Zhu}}, \bibinfo {author} {\bibfnamefont {H.}~\bibnamefont {Ramezani}},
  \bibinfo {author} {\bibfnamefont {C.}~\bibnamefont {Shi}}, \bibinfo {author}
  {\bibfnamefont {J.}~\bibnamefont {Zhu}},\ and\ \bibinfo {author}
  {\bibfnamefont {X.}~\bibnamefont {Zhang}},\ }\bibfield  {title} {\bibinfo
  {title} {Pt-symmetric acoustics},\ }\href@noop {} {\bibfield  {journal}
  {\bibinfo  {journal} {Physical Review X}\ }\textbf {\bibinfo {volume} {4}},\
  \bibinfo {pages} {031042} (\bibinfo {year} {2014})}\BibitemShut {NoStop}%
\bibitem [{\citenamefont {Christensen}\ \emph {et~al.}(2016)\citenamefont
  {Christensen}, \citenamefont {Willatzen}, \citenamefont {Velasco},\ and\
  \citenamefont {Lu}}]{christensen2016parity}%
  \BibitemOpen
  \bibfield  {author} {\bibinfo {author} {\bibfnamefont {J.}~\bibnamefont
  {Christensen}}, \bibinfo {author} {\bibfnamefont {M.}~\bibnamefont
  {Willatzen}}, \bibinfo {author} {\bibfnamefont {V.}~\bibnamefont {Velasco}},\
  and\ \bibinfo {author} {\bibfnamefont {M.-H.}\ \bibnamefont {Lu}},\
  }\bibfield  {title} {\bibinfo {title} {Parity-time synthetic phononic
  media},\ }\href@noop {} {\bibfield  {journal} {\bibinfo  {journal} {Physical
  Review Letters}\ }\textbf {\bibinfo {volume} {116}},\ \bibinfo {pages}
  {207601} (\bibinfo {year} {2016})}\BibitemShut {NoStop}%
\bibitem [{\citenamefont {Li}\ \emph {et~al.}(2019)\citenamefont {Li},
  \citenamefont {Rosendo-L{\'o}pez}, \citenamefont {Zhu}, \citenamefont {Fan},
  \citenamefont {Torrent}, \citenamefont {Liang}, \citenamefont {Cheng},
  \citenamefont {Christensen} \emph {et~al.}}]{li2019ultrathin}%
  \BibitemOpen
  \bibfield  {author} {\bibinfo {author} {\bibfnamefont {H.-x.}\ \bibnamefont
  {Li}}, \bibinfo {author} {\bibfnamefont {M.}~\bibnamefont
  {Rosendo-L{\'o}pez}}, \bibinfo {author} {\bibfnamefont {Y.-f.}\ \bibnamefont
  {Zhu}}, \bibinfo {author} {\bibfnamefont {X.-d.}\ \bibnamefont {Fan}},
  \bibinfo {author} {\bibfnamefont {D.}~\bibnamefont {Torrent}}, \bibinfo
  {author} {\bibfnamefont {B.}~\bibnamefont {Liang}}, \bibinfo {author}
  {\bibfnamefont {J.-c.}\ \bibnamefont {Cheng}}, \bibinfo {author}
  {\bibfnamefont {J.}~\bibnamefont {Christensen}}, \emph {et~al.},\ }\bibfield
  {title} {\bibinfo {title} {Ultrathin acoustic parity-time symmetric
  metasurface cloak},\ }\href@noop {} {\bibfield  {journal} {\bibinfo
  {journal} {Research}\ }\textbf {\bibinfo {volume} {2019}},\ \bibinfo {pages}
  {8345683} (\bibinfo {year} {2019})}\BibitemShut {NoStop}%
\bibitem [{\citenamefont {Fleury}\ \emph {et~al.}(2015)\citenamefont {Fleury},
  \citenamefont {Sounas},\ and\ \citenamefont {Alu}}]{fleury2015invisible}%
  \BibitemOpen
  \bibfield  {author} {\bibinfo {author} {\bibfnamefont {R.}~\bibnamefont
  {Fleury}}, \bibinfo {author} {\bibfnamefont {D.}~\bibnamefont {Sounas}},\
  and\ \bibinfo {author} {\bibfnamefont {A.}~\bibnamefont {Alu}},\ }\bibfield
  {title} {\bibinfo {title} {An invisible acoustic sensor based on parity-time
  symmetry},\ }\href@noop {} {\bibfield  {journal} {\bibinfo  {journal} {Nature
  Communications}\ }\textbf {\bibinfo {volume} {6}},\ \bibinfo {pages} {5905}
  (\bibinfo {year} {2015})}\BibitemShut {NoStop}%
\bibitem [{\citenamefont {Zhang}\ \emph {et~al.}(2018)\citenamefont {Zhang},
  \citenamefont {Peng}, \citenamefont {{\"O}zdemir}, \citenamefont {Pichler},
  \citenamefont {Krimer}, \citenamefont {Zhao}, \citenamefont {Nori},
  \citenamefont {Liu}, \citenamefont {Rotter},\ and\ \citenamefont
  {Yang}}]{zhang2018phonon}%
  \BibitemOpen
  \bibfield  {author} {\bibinfo {author} {\bibfnamefont {J.}~\bibnamefont
  {Zhang}}, \bibinfo {author} {\bibfnamefont {B.}~\bibnamefont {Peng}},
  \bibinfo {author} {\bibfnamefont {{\c{S}}.~K.}\ \bibnamefont {{\"O}zdemir}},
  \bibinfo {author} {\bibfnamefont {K.}~\bibnamefont {Pichler}}, \bibinfo
  {author} {\bibfnamefont {D.~O.}\ \bibnamefont {Krimer}}, \bibinfo {author}
  {\bibfnamefont {G.}~\bibnamefont {Zhao}}, \bibinfo {author} {\bibfnamefont
  {F.}~\bibnamefont {Nori}}, \bibinfo {author} {\bibfnamefont {Y.-x.}\
  \bibnamefont {Liu}}, \bibinfo {author} {\bibfnamefont {S.}~\bibnamefont
  {Rotter}},\ and\ \bibinfo {author} {\bibfnamefont {L.}~\bibnamefont {Yang}},\
  }\bibfield  {title} {\bibinfo {title} {A phonon laser operating at an
  exceptional point},\ }\href@noop {} {\bibfield  {journal} {\bibinfo
  {journal} {Nature Photonics}\ }\textbf {\bibinfo {volume} {12}},\ \bibinfo
  {pages} {479} (\bibinfo {year} {2018})}\BibitemShut {NoStop}%
\bibitem [{\citenamefont {Quan}\ \emph {et~al.}(2019)\citenamefont {Quan},
  \citenamefont {Sounas},\ and\ \citenamefont
  {Al{\`u}}}]{quan2019nonreciprocal}%
  \BibitemOpen
  \bibfield  {author} {\bibinfo {author} {\bibfnamefont {L.}~\bibnamefont
  {Quan}}, \bibinfo {author} {\bibfnamefont {D.~L.}\ \bibnamefont {Sounas}},\
  and\ \bibinfo {author} {\bibfnamefont {A.}~\bibnamefont {Al{\`u}}},\
  }\bibfield  {title} {\bibinfo {title} {Nonreciprocal willis coupling in
  zero-index moving media},\ }\href@noop {} {\bibfield  {journal} {\bibinfo
  {journal} {Physical Review Letters}\ }\textbf {\bibinfo {volume} {123}},\
  \bibinfo {pages} {064301} (\bibinfo {year} {2019})}\BibitemShut {NoStop}%
\bibitem [{\citenamefont {Vasseur}\ \emph {et~al.}(2007)\citenamefont
  {Vasseur}, \citenamefont {Hladky-Hennion}, \citenamefont {Djafari-Rouhani},
  \citenamefont {Duval}, \citenamefont {Dubus}, \citenamefont {Pennec},\ and\
  \citenamefont {Deymier}}]{vasseur2007waveguiding}%
  \BibitemOpen
  \bibfield  {author} {\bibinfo {author} {\bibfnamefont {J.}~\bibnamefont
  {Vasseur}}, \bibinfo {author} {\bibfnamefont {A.-C.}\ \bibnamefont
  {Hladky-Hennion}}, \bibinfo {author} {\bibfnamefont {B.}~\bibnamefont
  {Djafari-Rouhani}}, \bibinfo {author} {\bibfnamefont {F.}~\bibnamefont
  {Duval}}, \bibinfo {author} {\bibfnamefont {B.}~\bibnamefont {Dubus}},
  \bibinfo {author} {\bibfnamefont {Y.}~\bibnamefont {Pennec}},\ and\ \bibinfo
  {author} {\bibfnamefont {P.~A.}\ \bibnamefont {Deymier}},\ }\bibfield
  {title} {\bibinfo {title} {Waveguiding in two-dimensional piezoelectric
  phononic crystal plates},\ }\href@noop {} {\bibfield  {journal} {\bibinfo
  {journal} {Journal of applied physics}\ }\textbf {\bibinfo {volume} {101}},\
  \bibinfo {pages} {114904} (\bibinfo {year} {2007})}\BibitemShut {NoStop}%
\bibitem [{\citenamefont {Hladky-Hennion}\ and\ \citenamefont
  {Decarpigny}(1993)}]{hladky1993finite}%
  \BibitemOpen
  \bibfield  {author} {\bibinfo {author} {\bibfnamefont {A.-C.}\ \bibnamefont
  {Hladky-Hennion}}\ and\ \bibinfo {author} {\bibfnamefont {J.-N.}\
  \bibnamefont {Decarpigny}},\ }\bibfield  {title} {\bibinfo {title} {Finite
  element modeling of active periodic structures: Application to 1--3
  piezocompositesa},\ }\href@noop {} {\bibfield  {journal} {\bibinfo  {journal}
  {The Journal of the Acoustical Society of America}\ }\textbf {\bibinfo
  {volume} {94}},\ \bibinfo {pages} {621} (\bibinfo {year} {1993})}\BibitemShut
  {NoStop}%
\bibitem [{\citenamefont {Hou}\ and\ \citenamefont
  {Assouar}(2018)}]{hou2018tunable}%
  \BibitemOpen
  \bibfield  {author} {\bibinfo {author} {\bibfnamefont {Z.}~\bibnamefont
  {Hou}}\ and\ \bibinfo {author} {\bibfnamefont {B.}~\bibnamefont {Assouar}},\
  }\bibfield  {title} {\bibinfo {title} {Tunable elastic parity-time symmetric
  structure based on the shunted piezoelectric materials},\ }\href@noop {}
  {\bibfield  {journal} {\bibinfo  {journal} {Journal of Applied Physics}\
  }\textbf {\bibinfo {volume} {123}},\ \bibinfo {pages} {085101} (\bibinfo
  {year} {2018})}\BibitemShut {NoStop}%
\bibitem [{\citenamefont {Hou}\ \emph {et~al.}(2018)\citenamefont {Hou},
  \citenamefont {Ni},\ and\ \citenamefont {Assouar}}]{hou2018p}%
  \BibitemOpen
  \bibfield  {author} {\bibinfo {author} {\bibfnamefont {Z.}~\bibnamefont
  {Hou}}, \bibinfo {author} {\bibfnamefont {H.}~\bibnamefont {Ni}},\ and\
  \bibinfo {author} {\bibfnamefont {B.}~\bibnamefont {Assouar}},\ }\bibfield
  {title} {\bibinfo {title} {P t symmetry for elastic negative refraction},\
  }\href@noop {} {\bibfield  {journal} {\bibinfo  {journal} {Physical Review
  Applied}\ }\textbf {\bibinfo {volume} {10}},\ \bibinfo {pages} {044071}
  (\bibinfo {year} {2018})}\BibitemShut {NoStop}%
\bibitem [{\citenamefont {Wu}\ \emph {et~al.}(2019)\citenamefont {Wu},
  \citenamefont {Chen},\ and\ \citenamefont {Huang}}]{wu2019asymmetric}%
  \BibitemOpen
  \bibfield  {author} {\bibinfo {author} {\bibfnamefont {Q.}~\bibnamefont
  {Wu}}, \bibinfo {author} {\bibfnamefont {Y.}~\bibnamefont {Chen}},\ and\
  \bibinfo {author} {\bibfnamefont {G.}~\bibnamefont {Huang}},\ }\bibfield
  {title} {\bibinfo {title} {Asymmetric scattering of flexural waves in a
  parity-time symmetric metamaterial beam},\ }\href@noop {} {\bibfield
  {journal} {\bibinfo  {journal} {The Journal of the Acoustical Society of
  America}\ }\textbf {\bibinfo {volume} {146}},\ \bibinfo {pages} {850}
  (\bibinfo {year} {2019})}\BibitemShut {NoStop}%
\bibitem [{\citenamefont {Lu}\ and\ \citenamefont {Norris}(2020)}]{lu2020non}%
  \BibitemOpen
  \bibfield  {author} {\bibinfo {author} {\bibfnamefont {Z.}~\bibnamefont
  {Lu}}\ and\ \bibinfo {author} {\bibfnamefont {A.~N.}\ \bibnamefont
  {Norris}},\ }\bibfield  {title} {\bibinfo {title} {Non-reciprocal wave
  transmission in a bilinear spring-mass system},\ }\href@noop {} {\bibfield
  {journal} {\bibinfo  {journal} {Journal of Vibration and Acoustics}\ }\textbf
  {\bibinfo {volume} {142}},\ \bibinfo {pages} {021006} (\bibinfo {year}
  {2020})}\BibitemShut {NoStop}%
\bibitem [{SM()}]{SM}%
  \BibitemOpen
  \href@noop {} {}\bibinfo {howpublished} {See Supplemental Material at
  http://link.aps.org/supplemental/ for the derivation of the flexural wave
  equation modelling, detailed boundary conditions, equations and derivation of
  the transfer and scattering matrix formalism for flexural waves, as well as
  some results on the tunability of the EPs in the framework of flexural
  waves}\BibitemShut {NoStop}%
\bibitem [{bih()}]{biha}%
  \BibitemOpen
  \href@noop {} {}\bibinfo {howpublished} {$\beta$ is the flexural wavenumber,
  related to angular frequency through the quadratic dispersion
  $\beta^2=\omega\sqrt{\rho h/D}$, with $\rho$ the density of the TEP, $h$ its
  thickness, that is assumed to be small in comparison to the lateral
  dimensions of the TEP ($L$) as well as the wavelength, i.e. $h\ll L$ and
  $\beta h\ll 1$. $D$ denotes the flexural rigidity of the TEP, that is
  $Eh^3/[12(1-\nu^2)]$, with $E$ and $\nu$ being the Young's modulus and
  Poisson's ratio of the TEP, respectively.}\BibitemShut {Stop}%
\bibitem [{pla()}]{plate}%
  \BibitemOpen
  \href@noop {} {}\bibinfo {howpublished} {The width of these layers (in the
  $x$-direction) is assumed to be identical and is set as 20 cm. The thickness
  of the plate (in the $z$-direction) is identical for all layers and is set as
  2 cm. The other parameters of the plate are taken as follows: density, 2790
  $\textrm{kg}/\textrm{m}^3$, Poisson's ratio, 0.334, which correspond to a
  Duraluminum elastic plate.}\BibitemShut {Stop}%
\bibitem [{\citenamefont {Longhi}(2010)}]{longhi2010pt}%
  \BibitemOpen
  \bibfield  {author} {\bibinfo {author} {\bibfnamefont {S.}~\bibnamefont
  {Longhi}},\ }\bibfield  {title} {\bibinfo {title} {Pt-symmetric laser
  absorber},\ }\href@noop {} {\bibfield  {journal} {\bibinfo  {journal}
  {Physical Review A}\ }\textbf {\bibinfo {volume} {82}},\ \bibinfo {pages}
  {031801} (\bibinfo {year} {2010})}\BibitemShut {NoStop}%
\bibitem [{\citenamefont {Chong}\ \emph {et~al.}(2011)\citenamefont {Chong},
  \citenamefont {Ge},\ and\ \citenamefont {Stone}}]{chong2011p}%
  \BibitemOpen
  \bibfield  {author} {\bibinfo {author} {\bibfnamefont {Y.}~\bibnamefont
  {Chong}}, \bibinfo {author} {\bibfnamefont {L.}~\bibnamefont {Ge}},\ and\
  \bibinfo {author} {\bibfnamefont {A.~D.}\ \bibnamefont {Stone}},\ }\bibfield
  {title} {\bibinfo {title} {Pt-symmetry breaking and laser-absorber modes in
  optical scattering systems},\ }\href@noop {} {\bibfield  {journal} {\bibinfo
  {journal} {Physical Review Letters}\ }\textbf {\bibinfo {volume} {106}},\
  \bibinfo {pages} {093902} (\bibinfo {year} {2011})}\BibitemShut {NoStop}%
\bibitem [{\citenamefont {Sakhdari}\ \emph {et~al.}(2018)\citenamefont
  {Sakhdari}, \citenamefont {Estakhri}, \citenamefont {Bagci},\ and\
  \citenamefont {Chen}}]{sakhdari2018low}%
  \BibitemOpen
  \bibfield  {author} {\bibinfo {author} {\bibfnamefont {M.}~\bibnamefont
  {Sakhdari}}, \bibinfo {author} {\bibfnamefont {N.~M.}\ \bibnamefont
  {Estakhri}}, \bibinfo {author} {\bibfnamefont {H.}~\bibnamefont {Bagci}},\
  and\ \bibinfo {author} {\bibfnamefont {P.-Y.}\ \bibnamefont {Chen}},\
  }\bibfield  {title} {\bibinfo {title} {Low-threshold lasing and coherent
  perfect absorption in generalized pt-symmetric optical structures},\
  }\href@noop {} {\bibfield  {journal} {\bibinfo  {journal} {Physical Review
  Applied}\ }\textbf {\bibinfo {volume} {10}},\ \bibinfo {pages} {024030}
  (\bibinfo {year} {2018})}\BibitemShut {NoStop}%
\bibitem [{\citenamefont {Shi}\ \emph {et~al.}(2016)\citenamefont {Shi},
  \citenamefont {Dubois}, \citenamefont {Chen}, \citenamefont {Cheng},
  \citenamefont {Ramezani}, \citenamefont {Wang},\ and\ \citenamefont
  {Zhang}}]{shi2016accessing}%
  \BibitemOpen
  \bibfield  {author} {\bibinfo {author} {\bibfnamefont {C.}~\bibnamefont
  {Shi}}, \bibinfo {author} {\bibfnamefont {M.}~\bibnamefont {Dubois}},
  \bibinfo {author} {\bibfnamefont {Y.}~\bibnamefont {Chen}}, \bibinfo {author}
  {\bibfnamefont {L.}~\bibnamefont {Cheng}}, \bibinfo {author} {\bibfnamefont
  {H.}~\bibnamefont {Ramezani}}, \bibinfo {author} {\bibfnamefont
  {Y.}~\bibnamefont {Wang}},\ and\ \bibinfo {author} {\bibfnamefont
  {X.}~\bibnamefont {Zhang}},\ }\bibfield  {title} {\bibinfo {title} {Accessing
  the exceptional points of parity-time symmetric acoustics},\ }\href@noop {}
  {\bibfield  {journal} {\bibinfo  {journal} {Nature Communications}\ }\textbf
  {\bibinfo {volume} {7}},\ \bibinfo {pages} {1} (\bibinfo {year}
  {2016})}\BibitemShut {NoStop}%
\end{thebibliography}%

%%%%%%%%%%%%%%%%%%%%%%%%%%%%%%%%%%%%%%%Figures%%%%%%%%%%%%%%%%%%%%%%%%%%%%%%%%%%%%%%%%%%%%%%%%%%%%%%
%Figure
%%%%%%%%%%%%%%%%%%%%%%%%%%%%%%%%%%%%%%%%%%%%%%%%%%%%%%%%%%%%%%%%%%%%%%%%%%%%%%%%%%%%%%%%%%%%%%%%%

\end{document}